\documentclass[aip,amsmath,amssymb,
preprint,%
]{revtex4-1}

\usepackage{graphicx}
\usepackage{dcolumn}
\usepackage{bm}
\usepackage{soul}

\usepackage[utf8]{inputenc}
\usepackage[T1]{fontenc}
\usepackage{mathptmx}

\usepackage[colorlinks,linkcolor=blue,citecolor=blue,urlcolor=blue]{hyperref} 

\begin{document}

\title{Soft-wired long-term memory in a natural recurrent neuronal network}

\author{Miguel A. Casal}
\thanks{Equal contribution.}
\affiliation{Department of Experimental and Health Sciences, Universitat Pompeu Fabra, Barcelona Biomedical Research Park, Dr. Aiguader 88, 08003 Barcelona, Spain}
\affiliation{Istituto Italiano di Tecnologia, Center for Human Technologies, Via Enrico Melen 83, 16152 Genova, Italy}
\affiliation{Department of Computer Science, Bioengineering, Robotics and Systems Engineering (DIBRIS), University of Genova, Via Opera Pia 13, 16145 Genova}

\author{Santiago Galella}
\thanks{Equal contribution.}
\affiliation{Department of Experimental and Health Sciences, Universitat Pompeu Fabra, Barcelona Biomedical Research Park, Dr. Aiguader 88, 08003 Barcelona, Spain}

\author{Oscar Vilarroya}
\affiliation{
Department of Psychiatry and Legal Medicine, Universitat Aut\`onoma de Barcelona, Cerdanyola del Vall\`es 08193, Spain
}
\author{Jordi Garcia-Ojalvo}
\email{jordi.g.ojalvo@upf.edu}
\affiliation{Department of Experimental and Health Sciences, Universitat Pompeu Fabra, Barcelona Biomedical Research Park, Dr. Aiguader 88, 08003 Barcelona, Spain}


\begin{abstract}
Neuronal networks provide living organisms with the ability to process information.
They are also characterized by abundant recurrent connections, which give rise to strong feedback that dictates their dynamics and endows them with fading (short-term) memory.
The role of recurrence in \emph{long-term} memory, on the other hand, is still unclear.
Here we use the neuronal network of the roundworm \textit{C. elegans} to show that recurrent architectures in living organisms can exhibit long-term memory without relying on specific hard-wired modules.
A genetic algorithm reveals that the experimentally observed dynamics of the worm's neuronal network exhibits maximal complexity (as measured by permutation entropy).
In that complex regime, the response of the system to repeated presentations of a time-varying stimulus reveals a consistent behavior that can be interpreted as soft-wired long-term memory.
\end{abstract}


\maketitle

\begin{quotation}
A common manifestation of our ability to remember the past is the consistence of our responses to repeated presentations of stimuli across time.
Complex chaotic dynamics is known to produce such reliable responses in spite of its characteristic sensitive dependence on initial conditions.
In neuronal networks, complex behavior is known to result from a combination of (i) recurrent connections and (ii) a balance between excitation and inhibition.
Here we show that those features concur in the neuronal network of a living organism, namely {\em C. elegans}.
This enables long-term memory to arise in an on-line manner, without having to be hard-wired in the brain.
\end{quotation}

\section{Introduction
\label{sec:introduction}}

The nervous system of metazoans is composed of recurrent networks of neurons that allow them to respond to complex stimuli, both internal and external to the organism \cite{Douglas:1995aa,Sanchez-Vives:2000aa,Garrido:2007aa,Sancristobal:2016aa}.
Such recurrent architectures have inspired the design of specialized artificial neural networks \cite{Pearlmutter:1989aa,Graves:2013aa} that have revolutionized the field of machine learning, such as those based on long short-term memory (LSTM) circuits \cite{Hochreiter:1997aa,Sak:2014aa,Sainath:2015aa}. 
These systems rely on complex, hard-wired modules that provide them with memory capabilities, but which are too elaborate to have emerged naturally in living brains.

It thus seems reasonable to ask how the memory capabilities of living neuronal networks are implemented, and whether they require specific, hard-wired modules.
The recurrent character of biological circuits plays an important role here.
While the \emph{architecture} of recurrent neural networks endows them with the capacity to store static information \cite{Hopfield:1982aa}, their \emph{dynamics} provides them with fading memory \cite{Ganguli:2008aa} and with the ability to process time-dependent information in the short term, via a paradigm known as reservoir computing \cite{Maass:2002aa,Jaeger:2004aa,Appeltant:2011uq}.
This capability is not hard-wired into the system, but arises from the self-sustained dynamics provided by its recurrent connections: any external impulse perturbation will reverberate within the network for a while, mixing nonlinearly with its intrinsic dynamics
\cite{Buonomano:2009aa}.
This allows the on-line encoding of complex time-dependent inputs, which can be then decoded by a dedicated readout layer, located downstream of the recurrent core of the network (known as the reservoir) \cite{Lukosevicius2009}. The only connections whose weights need to be trained are those linking the reservoir with the readout layer, which underpins the effectiveness of this information-processing paradigm \cite{Verstraeten:2007aa}.

The standard reservoir computing (RC) architecture has not been widely adapted by the machine-learning community, nonetheless, because more sophisticated versions of it (such as the LSTM layout mentioned above) have proven more efficient with a bearable (for software-engineering standards) increase in complexity.
The simplicity of the basic RC structure, however, makes it still an attractive candidate for complex information processing in living organisms.
Here we use the soil nematode \textit{Caenorhabditis elegans} as model system to show that the architecture of living neuronal networks is compatible with the RC paradigm, and thereby displays fading (short-term) memory capabilities.
More importantly, we address the question of whether long-term memory can be similarly exhibited by such natural neuronal networks without specific, hard-wired memory modules.

\section{Topology of the \textit{C. elegans} neuronal network
\label{sec:topology}}

The architecture of the brains of higher animals is highly intricate \cite{Hagmann:2008aa,Bassett:2011aa} and variable \cite{Lange1997VariabilityYears}.
We thus chose a simpler organism for our study, namely the above-mentioned roundworm \textit{C. elegans}, whose neural system (in fact its entire cell lineage) is small and highly consistent from individual to individual. Additionally, and importantly for our purpose, the \textit{C. elegans} connectome has been fully mapped \cite{White1986TheElegans}.
From the approximately 1000 somatic cells forming the hermaphrodite worm, 302 compose its nervous system.
Those cells communicate to each other through around 6400 chemical synapses, 900 gap junctions and 1500 neuromuscular junctions \cite{White1986TheElegans}.

\begin{figure}[htb]
\centerline{\includegraphics[width=0.55\columnwidth]{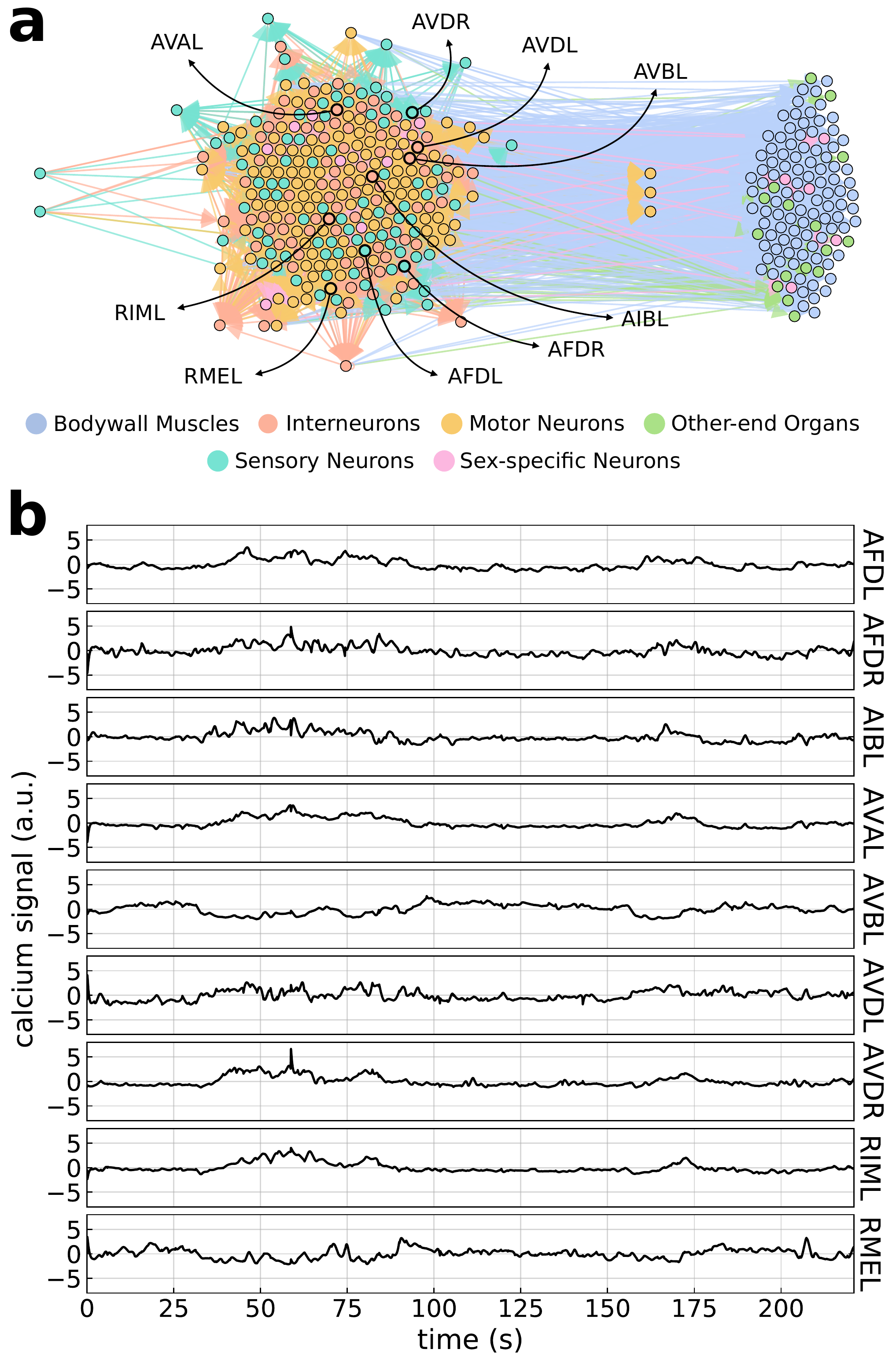}}
\caption{{\bf Topology and dynamics of the neuronal network of \textit{C. elegans}}.
(a) Cells listed in the connectome database of the worm \cite{Cook:2019aa} are represented as circles, colored according to the cell type (see legend), and connected by arrows indicating the presence of chemical coupling between them. The cells are clustered in such a way that the upstream input layer is located at the left of the plot (two sensory neurons on the far left), the downstream readout layer appears in the right, and the recurrent core is shown in the middle.
(b) Experimentally observed calcium signal denoting the dynamical activity of nine of the neurons of the reservoir, whose identities are highlighted in panel (a).
\label{fig:reservoir}}
\end{figure}

We used a recently published, updated connectivity map of \textit{C. elegans} \cite{Cook:2019aa}.
The data source provided the links between neurons and the associated weights (connection strengths), including not only canonical neurons, but also innervations from the neural system to muscles.
The dataset contained information of both electrical and chemical synapses, but here we focused only on the latter, which are considered more relevant for information-processing purposes.
The resulting graph contained two distinct components connected by only one link.
The smaller component contained only pharyngeal neurons, according to the WormAtlas database \cite{Altun2019WormAtlas}.
Given that the function of those neurons is very specific, we concentrated here on the largest component, which is shown in Fig.~\ref{fig:reservoir}a.
The nodes in the figure are colored according to the cell type (including muscle cells receiving innervations from neurons), and the links are colored according to the type of the receiving cell.

The next step was to identify the recurrent core of the network.
To do so, we pruned the graph \cite{Gabalda-Sagarra:2018ul} by iteratively removing nodes with either no outward connections (i.e. nodes that do not affect other nodes) or no inward connections (i.e. nodes that are not affected by other nodes).
The neurons preserved after this iterative procedure strictly belong to the reservoir, since any information they send out reaches back to them eventually.
The removed nodes, on the other hand, fall into two classes.
Those upstream of the recurrent core form the input layer (cf the two sensory neurons at the left of Fig.~\ref{fig:reservoir}a).
In turn, nodes downstream of the reservoir form the readout layer (cell clusters at the right of Fig.~\ref{fig:reservoir}a).
As expected, all non-neuronal cells in the network (those receiving innervations) belong to the readout layer.

Network pruning thus shows that the \textit{C. elegans} neuronal network displays the standard reservoir computing topology \cite{Verstraeten:2007aa,Lukosevicius2009}.
We now turn to analyzing the dynamics exhibited by this network.

\section{Network dynamics
\label{sec:dynamics}}

Advances in \textit{in vivo} calcium imaging have recently allowed monitoring the dynamical activity of tens of neurons in free-moving \textit{C. elegans} worms under controlled thermosensory conditions \cite{Venkatachalam2016Pan-neuronalElegans}.
We aimed to use such data to constrain the dynamics of the network established in Sec.~\ref{sec:topology} above.
Nine of the monitored neurons could be identified with cells listed in the connectome database analyzed above \cite{Cook:2019aa}.
The dynamics of those neurons, measured in terms of the time-resolved calcium signal in each cell, is shown in Fig.~\ref{fig:reservoir}b.
The neurons exhibited irregular waveforms, a dynamical trait that is known to result from a balance between excitation and inhibition \cite{vanVreeswijk1996ChaosActivity}.
Unfortunately we could not corroborate this fact with the connectome data used in the previous section \cite{Cook:2019aa}, since the database contained no systematic information on the excitatory/inhibitory character of the connections.
This issue, together with our inability to identify most of the neurons of which we had calcium signaling
data, made us pursue a parameter inference approach using a genetic algorithm on a dynamical model of the neuronal network.

To simulate the dynamics of our network, we used a standard discrete-time model in which the activity of the each neuron $i$ depends in a sigmoidal (threshold-like) manner on the inputs coming from other neurons:
\begin{equation}
x_{i,t+1} = \tanh \bigg ( \sum_j \omega_{ij}x_{j,t}\bigg)
\label{eq:update}
\end{equation}
where $x_{i,t}$ is the state of neuron $i$ at time $t$ and $\omega_{ij}$ represents the strength of the connection from neuron $j$ to neuron $i$, normalized such that the maximum connection strength in the network is 1.
The parameters $\omega_{ij}$ are positive for excitatory connections and negative for inhibitory ones.

Since the calcium monitoring experiments were performed under conditions of oscillating temperature, and two of the nine labeled neurons (AFDL and AFDR) are associated with thermotaxis \cite{Mori1995NeuralElegans}, we considered the experimentally measured activities of those two neurons (normalized between $+1$ and $-1$) as inputs to the network model (ignoring incoming connections acting upon them).
The activity of all other neurons is computed according to model (\ref{eq:update}).
The fitness of the $\{\omega_{ij}\}$ parameter set is then evaluated as the mean Pearson correlation coefficient between the experimental and modeled time traces of all seven remaining neurons in Fig.~\ref{fig:reservoir}b.

\begin{figure}[htb]
\centerline{\includegraphics[width=0.55\columnwidth]{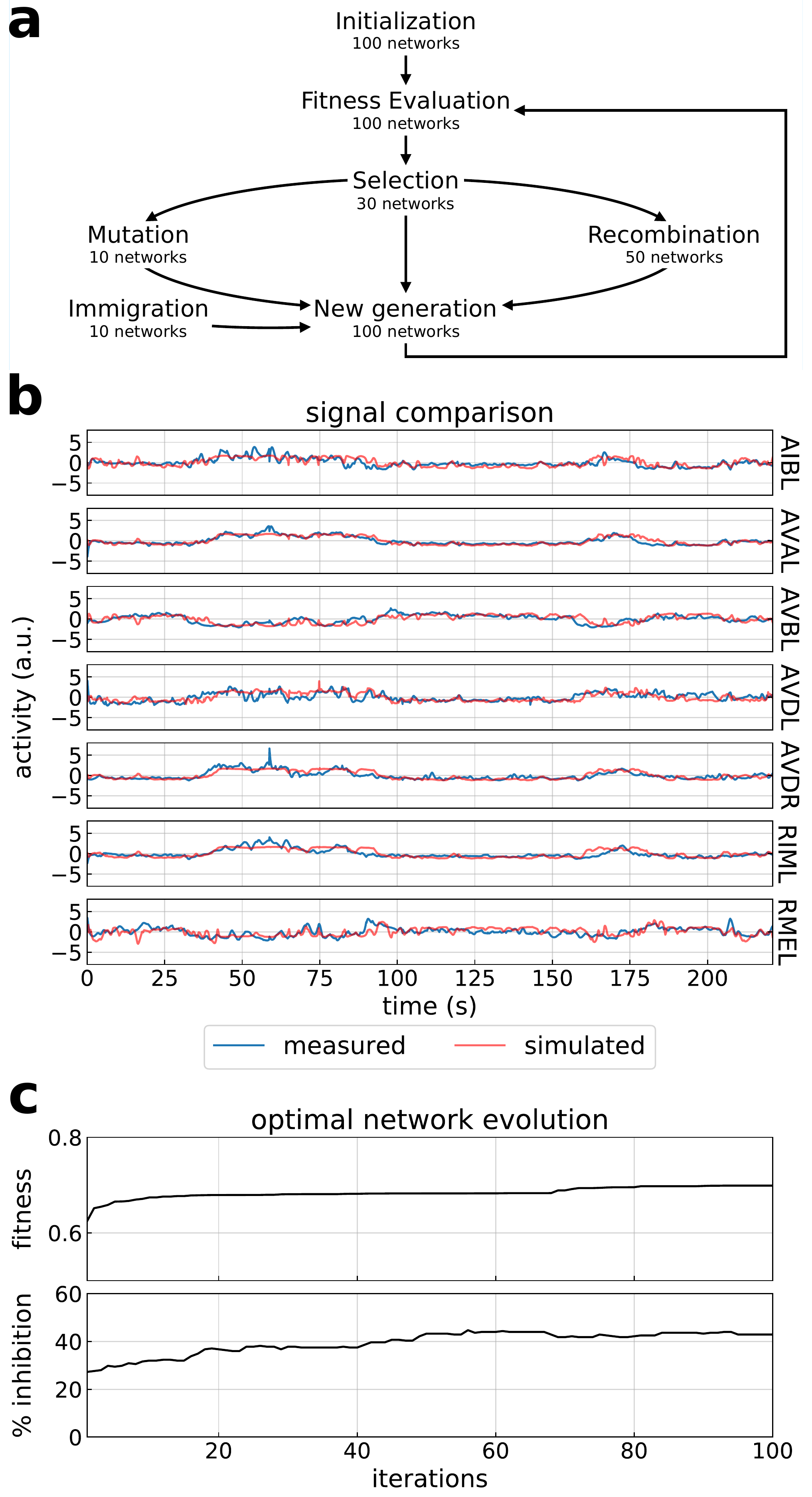}}
\caption{\textbf{Statistical inference of the connection types}.
(a) Scheme of the genetic algorithm used. 
(b) Dynamics of the seven non-input neurons generated by a particular instance of the model (blue lines) compared with the experimental data (red lines).
(c) Evolution of the fitness of the optimal individual network at each iteration of the genetic algorithm (top), and corresponding percentage of inhibitory connections (bottom).
\label{fig:genalg}}
\end{figure}

To maximize the fitness of our network model, the genetic algorithm (Fig.~\ref{fig:genalg}a) begins by choosing random signs for the connection strengths between all neurons in the network
(selecting a uniformly distributed random number of neurons from the network and making them inhibitory, with the rest being excitatory), while their magnitudes are fixed by the values given in the database.
A population of 100 networks built in this manner is modeled as described above.
After the dynamics is generated using Eq.~(\ref{eq:update}), we evaluate the fitness for each individual network.
Next, the 30 individuals with highest fitness were selected and used to generate an offspring of 50 children by recombining the adjacency matrices of randomly generated pairs of such individuals. 
Recombination was performed by selecting a cutoff neuron randomly, splitting the rows of the adjacency matrix at that neuron, and the rows above it from one parent and below it from the other.
This results in a population of 80 individuals (pooling the 30 parents and the 50 children).
The remaining 20 individuals were replaced using mutation and immigration.
Mutation consisted in selecting the best individual network and changing the signs of 5 connections chosen at random.
We repeated this process until having 10 different mutant individuals. The last 10 individuals were introducing through immigration.
These individuals were created with a random distribution of excitation and inhibition, as in the initial individuals of the algorithm described above (to avoid biasing the process).
The full set of new 100 individuals obtained is then used to start a new iteration cycle.
The dynamics of the optimal model resulting from the procedure described above is shown in Fig.~\ref{fig:genalg}b, where it is compared with the experimental observations.
Figure~\ref{fig:genalg}c shows the evolution of the model with highest fitness as the genetic algorithm was iterated.

\begin{figure}[htb]
\centerline{\includegraphics[width=0.65\columnwidth]{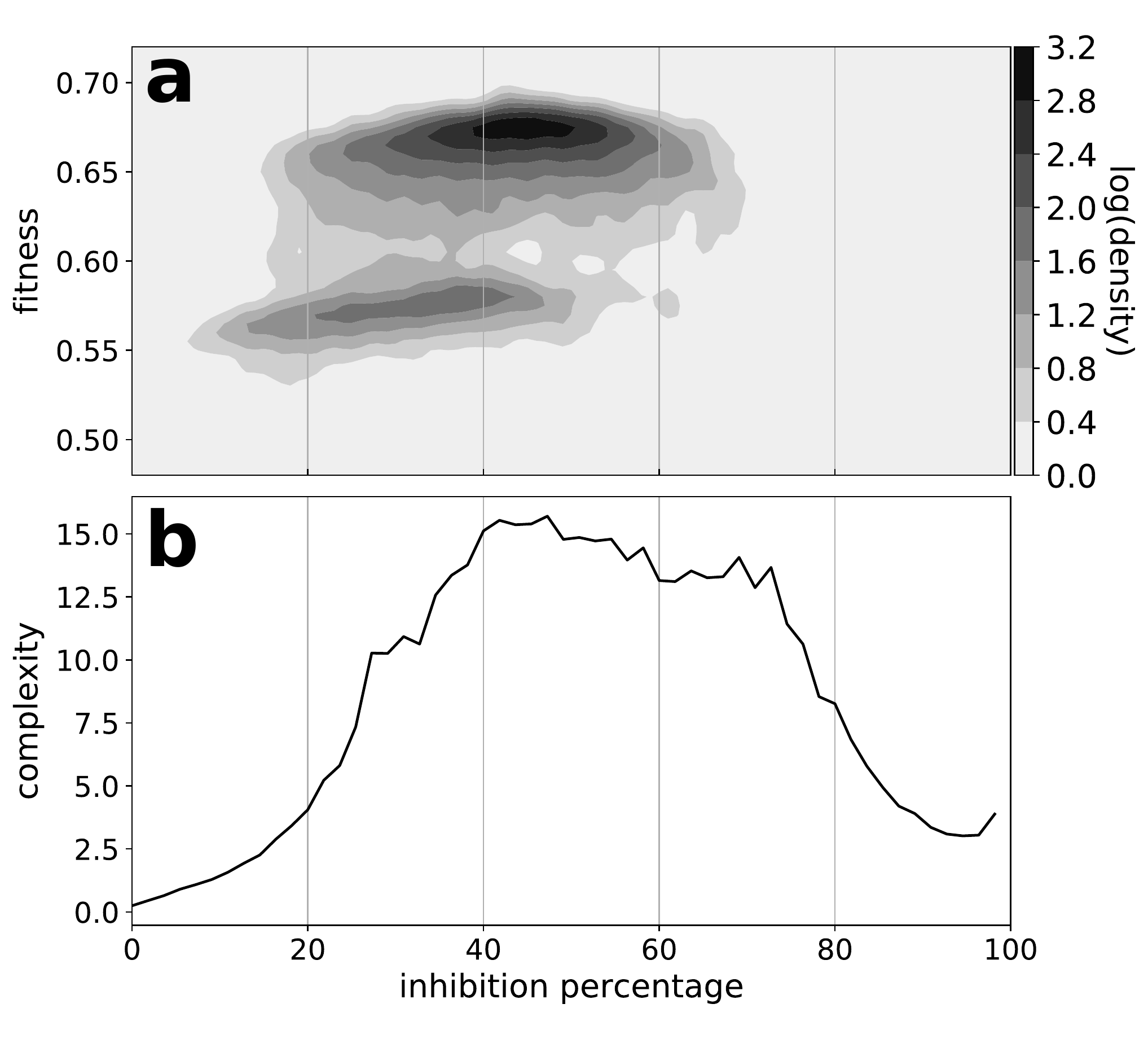}}
\caption{\textbf{Fitting to experiments reveals excitation-inhibition balance and complex dynamics.}
(a) Density plot of the fitness versus inhibition percentage, computed at each of 100 iterations for 1000 realizations of the genetic algorithm described in Fig.~\ref{fig:genalg}.
The density was smoothed out with a Gaussian filter.
Black corresponds to the highest density of optimal individuals.
(b) Complexity of the dynamics of the network for each inhibition percentage.
\label{fig:landscape}}
\end{figure}

In order to explore systematically the fitness landscape of our network model, we run the genetic algorithm described above for 100 iterations and 1000 realizations, starting from a corresponding number of adjacency matrices with a random balance between excitation and inhibition as described above, and identified the optimal network at each iteration, resulting in a total amount of $\sim 10^5$ individuals.
We then computed the fitness of the optimal individuals in each case, as a function of the percentage of inhibition.
The corresponding density distribution is shown in Fig.~\ref{fig:landscape}a, and indicates that the majority of optimal individuals are also those with highest fitness, and have an inhibition percentage in the range of 38\% to 55\%, approximately.
The existence of a balanced degree of inhibition and excitation is consistent with what is usually found in the brains of higher animals \cite{Yizhar2011pp171,malagarriga}, and is also in line with the complex dynamics reported above \cite{vanVreeswijk1996ChaosActivity}.

We also quantified the degree of complexity of the dynamical behavior using a standard measure of complexity, namely the permutation entropy \cite{bandt}. 
To do so, different we generated 1000 individuals for each inhibition percentage and calculated the permutation entropy for each, computing the average and standard deviation as a function of the inhibition percentage.
The complexity of the dynamics is quantified as the average of the permutation entropy divided by its standard deviation.
The dependence of the resulting complexity coefficient on the inhibition percentage is shown in Fig.~\ref{fig:landscape}b.
The result confirms that the complexity of the dynamics is high when the network model optimally represents the experimental observations.

\section{Response of the network to a pulse-train stimulation
\label{sec:stimulation}}

The recurrent architecture of the \textit{C. elegans} neuronal network revealed in Sec.~\ref{sec:topology} endows the system with short-term memory.
We conjectured that the complex dynamics shown in the previous section, on the other hand, makes it possible for the network to exhibit long-term memory as well.
To test this hypothesis, we first examine the response of the network to external stimulation.
This can be modeled in our neuronal network by modifying the update rule (\ref{eq:update}) as follows:
\begin{equation}
x_{i,t+1} = \tanh \bigg( v_iz_t + \sum_j \omega_{ij}x_{j,t}\bigg)%
\end{equation}
Here $z_t$ is the external input at time $t$ and $v_i$ is the weight between the input and neuron $i$.
In this model, the transfer function (hyperbolic tangent) integrates the state of the neurons
in each time step with the external input that they receive. We chose the weights $v_i$ randomly for each neuron, so that the input affects more strongly some neurons than others.
The initial condition is chosen randomly for each neuron, uniformly distributed between $-1$ and $1$.

\begin{figure}
\centerline{\includegraphics[width=0.55\columnwidth]{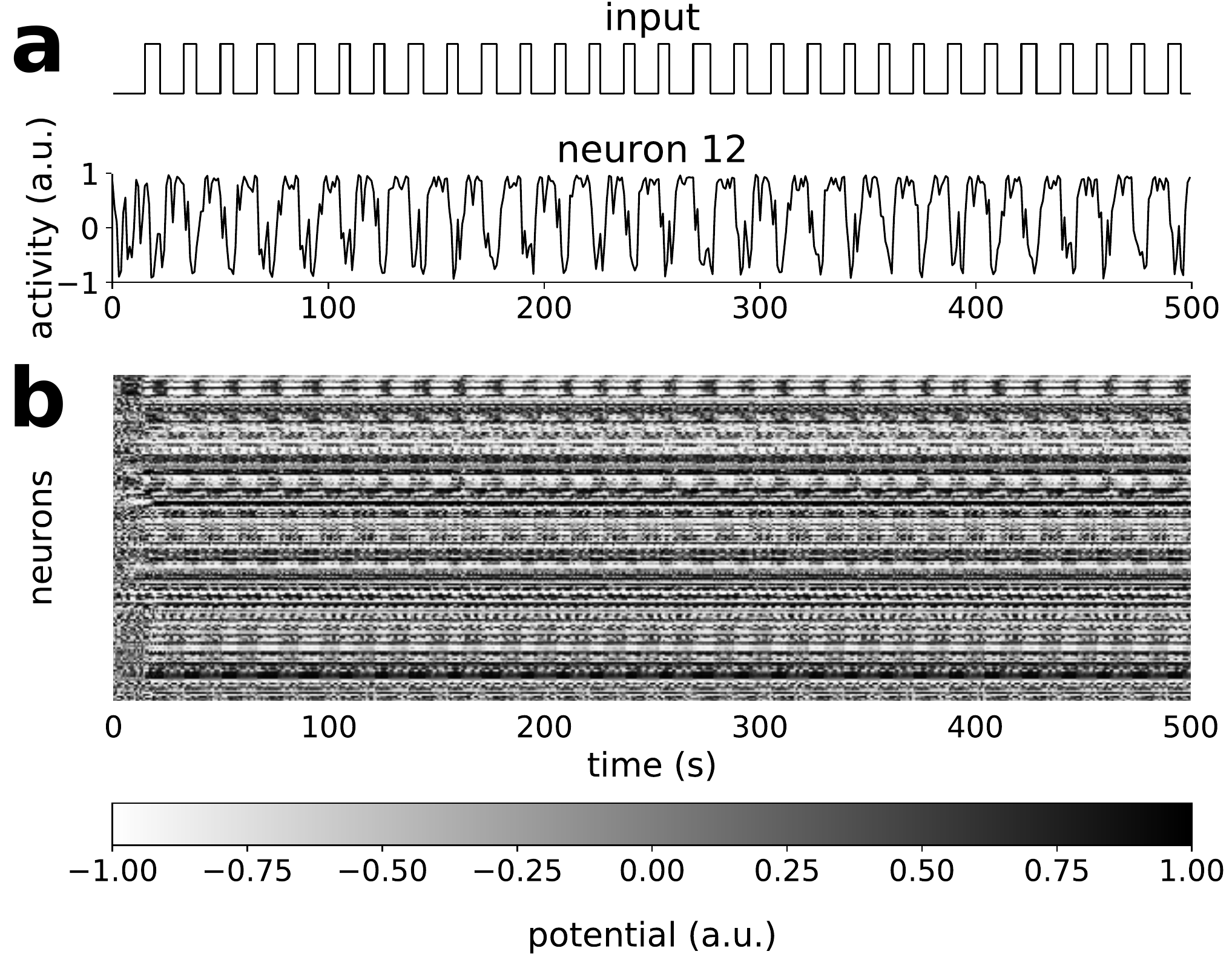}}
\caption{\textbf{Response of the \textit{C. elegans} network to pulsatile stimulation.}
(a) The top panel shows the input applied to the network, while the bottom panel represents the response of one of the neurons (neuron 12, ASER).
(b) Response of the entire network to the input signal shown at the top of panel (a), with the state of each neuron represented in gray according with the scale bar at the bottom.
\label{fig:response}}
\end{figure}

The typical response of the network to a train of square pulses alternating between two values (considered binary from simplicity, 0 or 1) is shown in Fig.~\ref{fig:response}.
The duration of the 0 state is chosen constant and equal to 10 time steps, whereas that of the 1 state is randomly chosen uniformly in the interval $8\pm 2$ time units.
The top plot in Fig.~\ref{fig:response}a shows the input, and the bottom plot in that panel displays the dynamics of a given neuron of the network (taken to be neuron 12, ASER, in what follows).
In turn, Fig.~\ref{fig:response}b shows the dynamics of all the neurons in the network, represented in grayscale from $-1$ to $1$.
The figure shows that the stimulation maintains the system continuously in transient chaos (we note that the model contains no noise source), with each change in the input (from 0 to 1 and vice versa) eliciting a chaotic relaxational dynamics that does not repeat from pulse to pulse.

\begin{figure}
\centerline{\includegraphics[width=0.55\columnwidth]{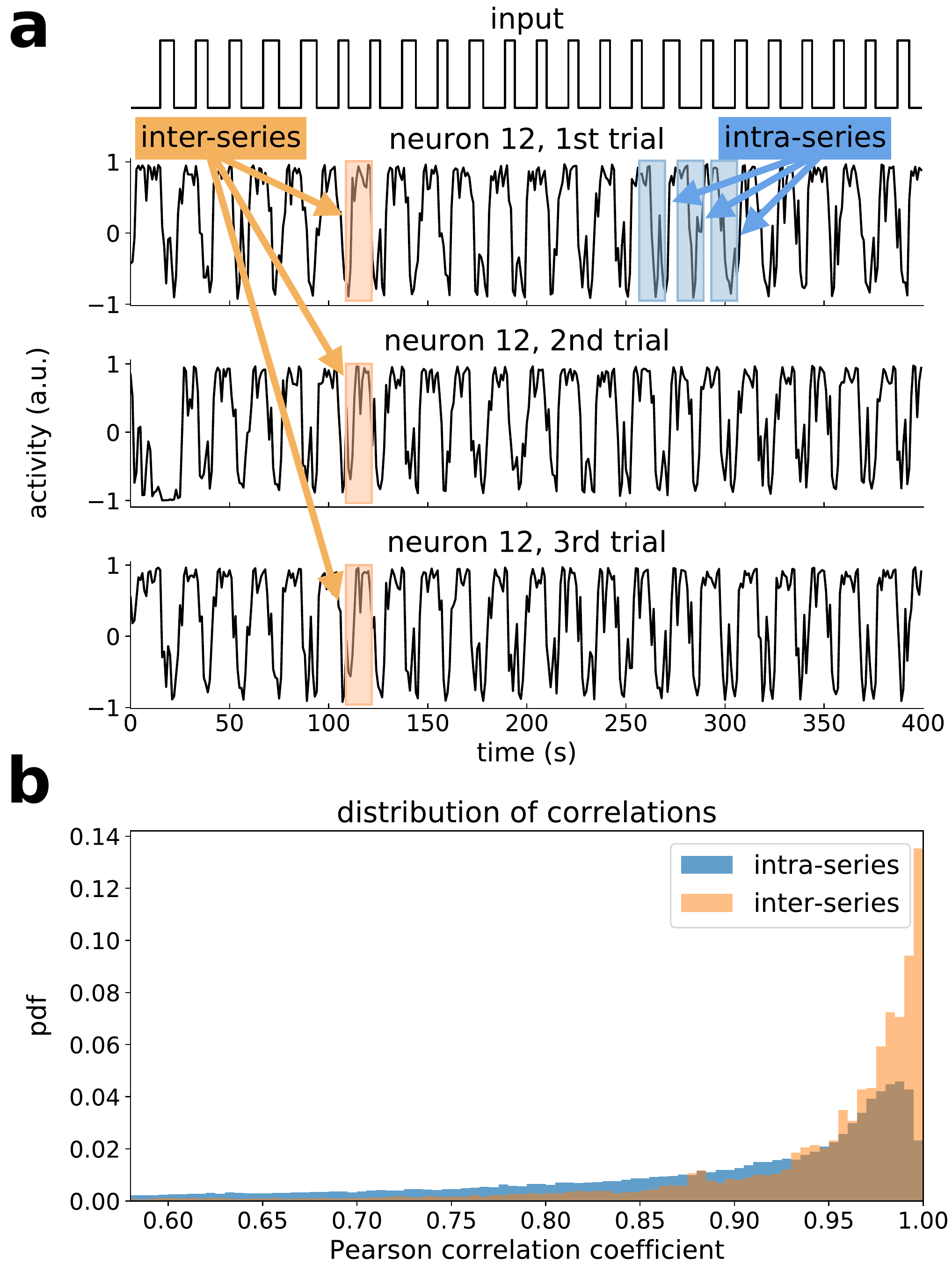}}
\caption{\textbf{Response of the \textit{C. elegans} network to repeated irregular stimulation.}
(a) Response of neuron 12 (bottom three panels) to three presentations of the same input (top panel).
(b) Distribution of the correlation coefficient between pairs of responses to repeated presentations of the same signal across trials (inter-series comparison, orange) and between pulses along time in the same trial (intra-series, blue). $10^5$ pairs of responses were compared in each case.
\label{fig:repeated}}
\end{figure}
\begin{figure}[htb]
\centerline{\includegraphics[width=0.9\columnwidth]{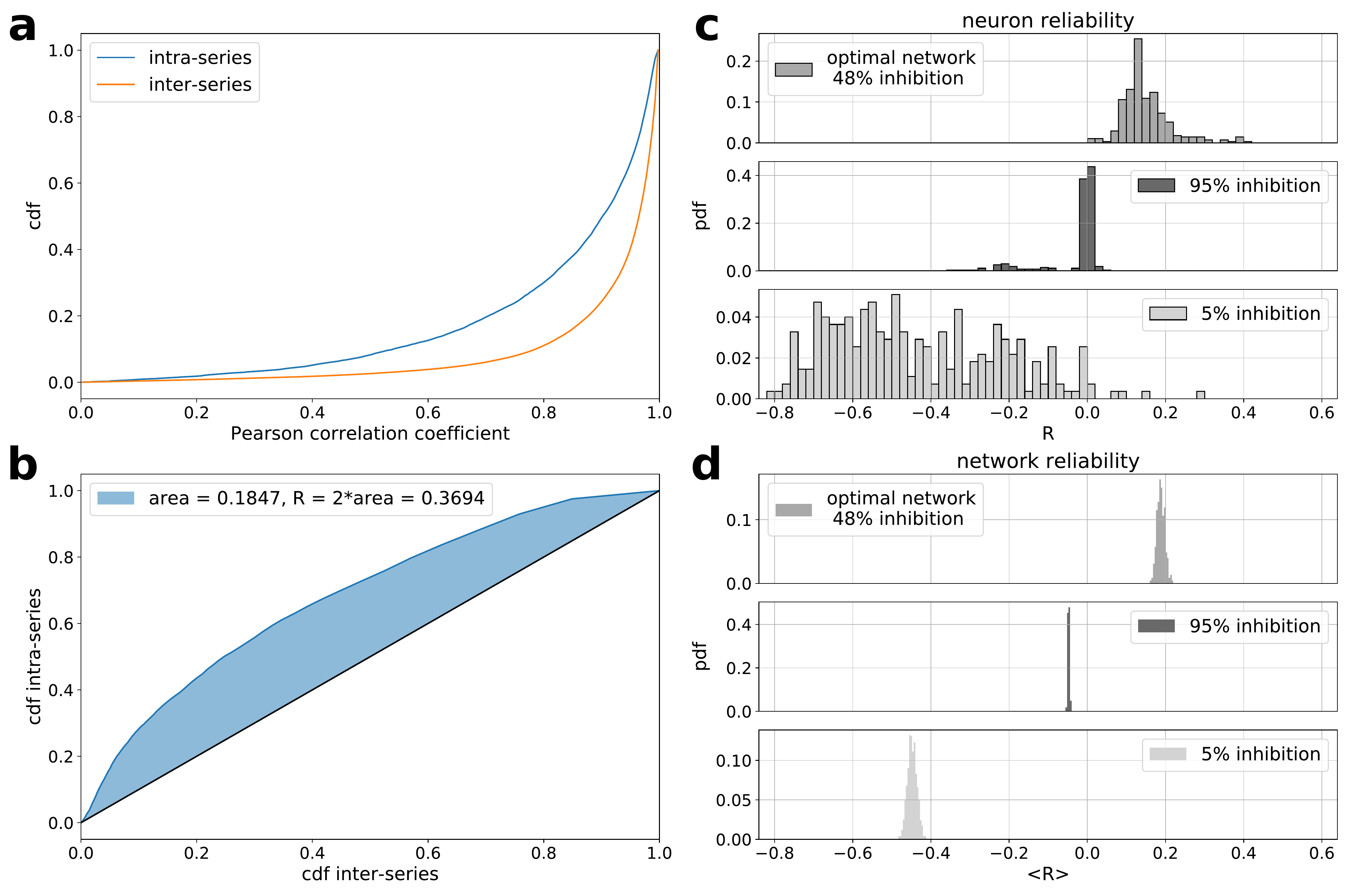}}
\caption{\textbf{Quantifying the reliable response of the \textit{C. elegans} network.}
(a) Cumulative distribution function of the Pearson correlation coefficient between intra-series (blue) and inter-series (orange) pairs, for neuron 12 (Fig.~\ref{fig:repeated}) in a network with 48\% inhibition.
(b) Corresponding receiver operating characteristic (ROC) curve of the two cumulative distribution functions, for the two distributions shown in panel a.
(c) Distribution of the reliability coefficient for individual neurons, defined in terms of the blue area of panel b (see text).
(d) Distribution of the reliability coefficient averaged over neurons for different network realizations.
\label{fig:cdf}}
\end{figure}

\section{Response to repeated stimulations
\label{sec:repeat}}

When next applied repeated presentations of the same irregular pulsed stimulus, with the goal of establishing whether the response of the network is consistent between trials.
We interpret a consistent response in terms of long-term memory.
As a reference, we compared the response of the network to the different trials with the response to subsequent pulses along time (Fig.~\ref{fig:repeated}a).
Specifically, we applied 20 consecutive pulses to the network, focusing on its response to the drop from 1 to 0.
We compared the response to each of the 20 inputs between one reference series and 5000 additional trials (inter-series comparison, orange intervals in Fig.~\ref{fig:repeated}a).
We also compared the response of the network between all pairs of 20 events in the same trial along time (intra-series comparison, blue intervals in Fig.~\ref{fig:repeated}a), for a number of trials (526) such that in both cases the number of pairs was similar, $\sim 10^5$.
Quantifying the similarity between the response pairs in terms of the Pearson correlation coefficient (Fig.~\ref{fig:repeated}b), we found significant differences between the distributions of these coefficients in the two cases, with the inter-series correlation approaching 1 more frequently than the intra-series one.
Thus, the network responds more consistently to repeated presentations of the stimulus (change in the input from 1 to 0) across trials than across time.

To quantify systematically this reliability, we developed a quantifier based on the cumulative correlation distributions shown in Fig.~\ref{fig:cdf}a (obtained directly from Fig.~\ref{fig:repeated}b).
A reliable response is reflected in a later increase towards 1 in the cumulative distribution for the inter-series correlation coefficients.
Thus the fact that the cumulative distribution increases significantly earlier for the intra-series (blue) than for the inter-series (orange) responses is an indication of the consistent behavior of the system.
Such difference can be quantified systematically with the receiver-operating curve (ROC) shown in Fig.~\ref{fig:cdf}b.
The area between that curve and the 45-degree line is a good estimator of the degree of reliability of the network: for equal responses across trials and across time (no reliability), the two cumulative distributions would be identical and the area would be zero; for perfect reliability, on the other hand, the area would be that of the full triangle above the 45-degree line ($1/2$).
We thus define the degree of reliability $R$ as the value of that area divided by its maximum value.
In the particular case of neuron 12 (Fig.~\ref{fig:cdf}b), that quantity is $\sim 0.37$.

Figure~\ref{fig:cdf}c shows the distribution of $R$ values for all neurons in the network, for three cases: the optimal network identified with the fitting above (top panel), and two control cases corresponding to excessive inhibition (middle panel) and excessive excitation (bottom panel).
The figure shows that in the case of the optimal fit to the experimental data (balanced excitation and inhibition) reliability is largest, being close to zero for 95\% inhibition, and broad and mostly negative for 5\% inhibition (a negative value of $R$ corresponds to a mostly periodic --and thus non-complex-- dynamics, which can be expected to be elicited in the network when excitation dominates).
Averaging $R$ over all neurons in the network for 1000 realizations of the stimulus (and initial conditions) leads to the distributions shown in Fig.~\ref{fig:cdf}d, which evidence the reproducibility of the results and the difference in reliability between the three types of network.

\section{Discussion
\label{sec:conclusion}}

The results shown above indicate that the neuronal network of a living organism can respond consistently to repeated presentations of a complex (irregular) sequence of events, even though the individual responses to single events across the sequence are highly variable.
In other words, complex series of stimuli trigger the same dynamical trajectory in the high-dimensional phase space of the neuronal network, when presented repeatedly in time.
This behavior builds upon the ability of the brain the encode information in its transient dynamics \cite{Rabinovich:2008fk}.
The resulting reliability can be interpreted as a form of a long-term memory that is encoded in the repertoire of dynamical attractors of the neuronal network, and can thus be considered soft-wired rather than hard-wired in the brain.

A key component in the behavior reported above is the existence of widespread feedback in the underlying neuronal network.
As we have shown in Sec.~\ref{sec:topology} above, a sizable fraction of the connections in the neuronal system of \textit{C. elegans} form cycles, and are therefore involved in internal feedback loops.
Additionally, fitting the \emph{in silico} behavior of the network to the experimentally measured dynamics of a subset of neurons reveals (Sec.~\ref{sec:dynamics}) an approximately balanced degree of excitation (positive connections) and inhibition (negative connections).
Neuronal networks with balanced excitation and inhibition are well known to produce self-sustained chaotic dynamics \cite{vanVreeswijk1996ChaosActivity}.
And chaotic systems have been reported to respond consistently to complex external signals: upon repeated presentation of these signals, a chaotic system is able to respond always in the same manner, in spite of the well known sensitive dependence on initial conditions that characterizes chaos.
This type of behavior was originally termed ``generalized synchronization'' \cite{Rulkov:1995aa,Abarbanel:1996aa,Kocarev:1996aa}, and has since been reported in a variety of physical systems including mechanical oscillators \cite{Tang:1998aa}, lasers \cite{Uchida:2003aa,McAllister:2004aa}, and spatial light modulators \cite{Uchida:2004aa}.
Here we have extended this phenomenology to living systems, specifically to neuronal networks within the context of reservoir computing. We interpret the resulting consistent dynamics in terms of soft-wired (as opposed to hard-wired) long-term memory.

The dynamics of our experimentally constrained neuronal network was generated using a model widely employed in studies of artificial neural networks.
This model captures the main information-processing features of more detailed biophysical models such as the the integrate-and-fire \cite{Abbott1999Lapicques1907} and Hodgkin-Huxley \cite{Hodgkin1952} models, and thus we expect the same behavior to be reproduced by those models.
A second limitation of our study is the choice of external input considered (a binary input, with the random duration of one of the two states as the only source of irregularity).
This choice was dictated by our need to quantify the consistency of the system response with respect to a reference situation (taken here to be the response to sequential input changes comparable to those being compared across trials).
It would be worthwhile to explore the response of the network to more complex inputs.
In any case, we believe that our study points to information-processing capabilities of biological networks that go beyond the paradigms considered so far.

\section*{Acknowledgments}
We thank Vivek Venkatachalam for kindly providing us with the experimental measurements of neuronal activity in moving {\em C. elegans} worms.
This work was supported by the Spanish Ministry of Science, Innovation and Universities and FEDER (project PGC2018-101251-B-I00 and ``Maria de Maeztu'' Programme for Units of Excellence in R\&D, grant CEX2018-000792-M), and by the Generalitat de Catalunya (ICREA Academia programme).
M.A.C. is currently supported by the EU Marie Sk\l{}odowska-Curie Training Network ``NeuTouch'' (contract 813713, call H2020-MSCA-ITN-2018).

\end{document}